\documentclass[final,nofootinbib,aps,twocolumn,showpacs,showkeys,groupaddress,preprintnumbers,floatfix]{revtex4-1}

\usepackage{dynlearn}

\newcommand{\source} [1] {\ensuremath{X_{#1}}\xspace}
\newcommand{\target}     {\ensuremath{Y}\xspace}

\newcommand{\sep}{\cdot}

\newcommand{\PID} [3] {
  \ensuremath{
    \operatorname{I_{#1}}
    \if\relax\detokenize{#3}\relax
      \!\!
    \else
      \left[ #2 \rightarrow #3 \right]
    \fi
  }
  \xspace
}

\newcommand{\Ipart}  [2] [\target] {\PID{\partial}{#2}{#1}}

\newcommand{\Ibroja} [2] [\target] {\PID{\textrm{BROJA}}{#2}{#1}}

\newcommand{\Idep}   [2] [\target] {\PID{\textrm{dep}}{#2}{#1}}

\newcommand{\pbar}{\ensuremath{\overline{p}}\xspace}

\newcommand{\SKARzero} [3] {\ensuremath{\operatorname{S}(#1 : #2 ~||~ #3)}\xspace}
\newcommand{\SKARonel} [3] {\ensuremath{\operatorname{S}(#1 \leftarrow #2 ~||~ #3)}\xspace}
\newcommand{\SKARoner} [3] {\ensuremath{\operatorname{S}(#1 \rightarrow #2 ~||~ #3)}\xspace}
\newcommand{\SKARone}  [3] {\SKARoner{#1}{#2}{#3}}
\newcommand{\SKARtwo}  [3] {\ensuremath{\operatorname{S}(#1 \leftrightarrow #2 ~||~ #3)}\xspace}

\usepackage{pifont}

\newcommand{\xmark}{\textcolor{red}{\ding{55}}}

\newcommand{\assumption}{\textbf{(\textasteriskcentered)}\xspace}

\usepackage{textcomp}
 
\begin{document}

\def\ourTitle{
  Unique Information and Secret Key Agreement
}

\def\ourAbstract{
The \emph{partial information decomposition} (PID) is a promising framework for decomposing a joint random variable into the amount of influence each source variable \source{i} has on a target variable \target, relative to the other sources.
For two sources, influence breaks down into the information that both \source{0} and \source{1} \emph{redundantly} share with \target, what \source{0} \emph{uniquely} shares with \target, what \source{1} \emph{uniquely} shares with \target, and finally what \source{0} and \source{1} \emph{synergistically} share with \target.
Unfortunately, considerable disagreement has arisen as to how these four components should be quantified.
Drawing from cryptography, we consider the \emph{secret key agreement rate} as an operational method of quantifying unique informations.
Secret key agreement rate comes in several forms, depending upon which parties are permitted to communicate.
We demonstrate that three of these four forms are inconsistent with the PID.
The remaining form implies certain interpretations as to the PID's meaning---interpretations not present in PID's definition but that, we argue, need to be explicit.
These reveal an inconsistency between third-order connected information, two-way secret key agreement rate, and synergy.
Similar difficulties arise with a popular PID measure in light the results here as well as from a maximum entropy viewpoint.
We close by reviewing the challenges facing the PID.
}

\def\ourKeywords{
  information theory, partial information decomposition, secret key agreement, cryptography
}

\hypersetup{
  pdfauthor={James P. Crutchfield},
  pdftitle={\ourTitle},
  pdfsubject={\ourAbstract},
  pdfkeywords={\ourKeywords},
  pdfproducer={},
  pdfcreator={}
}

\author{Ryan G. James}
\email{rgjames@ucdavis.edu}

\author{Jeffrey Emenheiser}
\email{jemenheiser@ucdavis.edu}

\author{James P. Crutchfield}
\email{chaos@ucdavis.edu}

\affiliation{Complexity Sciences Center and Physics Department,
University of California at Davis, One Shields Avenue, Davis, CA 95616}

\date{\today}
\bibliographystyle{unsrt}

\title{\ourTitle}

\begin{abstract}
\ourAbstract
\end{abstract}



\preprint{\arxiv{1811.XXXX}}

\title{\ourTitle}
\date{\today}
\maketitle

\setstretch{1.1}


\section{Introduction}
\label{sec:introduction}

Consider a joint distribution over ``source'' variables \source{0} and \source{1} and ``target'' \target.
Such distributions arise in many settings: sensory integration, logical computing, neural coding, functional network inference, and many others.
One promising approach to understanding how the information shared between $\left(\source{0}, \source{1}\right)$, and \target is organized is the \emph{partial information decomposition} (PID)~\cite{williams2010nonnegative}.
This decomposition seeks to quantify how much of the information shared between \source{0}, \source{1}, and \target is done so \emph{redundantly}, how much is \emph{uniquely} attributable to \source{0}, how much is \emph{uniquely} attributable to \source{1}, and finally how much arises \emph{synergistically} by considering both \source{0} and \source{1} together.

Unfortunately, the lack of a commonly accepted method of quantifying these components has hindered PID's adoption.
In point of fact, several proposed axioms are not mutually consistent~\cite{rauh2017extractable,rauh2017secret}.
And, to date, there is little agreement as to which should hold.
Here, we take a step toward understanding these issues by adopting an operational definition for the unique information.
This operational definition comes from information-theoretic cryptography and quantifies the rate at which two parties can construct a secret while a third party eavesdrops.

There are four varieties of secret key agreement rate depending on which parties are allowed to communicate, each of which defines a different PID.
Each variety also relates to a different intuition as to how the PID operates.
We discuss several aspects of these different methods and further demonstrate that three of the four fail to construct an internally consistent decomposition.

Our development proceeds as follows.
Section\nobreakspace \ref {sec:pid} briefly describes the two-source PID.
Section\nobreakspace \ref {sec:secrets} reviews the notion of secret key agreement rate and how to quantify it in three contexts: No one communicates, only Alice communicates, and both Alice and Bob communicate.
Section\nobreakspace \ref {sec:decompositions} discusses the behavior of the PID quantified utilizing secret key agreement rates as unique informations and what intuitions are implied by the choice of who is permitted to communicate.
Section\nobreakspace \ref {sec:discussion} explores two further implications of our primary results, first in a distribution where two-way communication seems to capture synergistic, third-order connected information and second in the behavior of an extant method of quantifying the PID along with maximum entropy methods.
Finally, Section\nobreakspace \ref {sec:conclusion} summarizes our findings and speculates about PID's future.

\section{Partial Information Decomposition}
\label{sec:pid}

Two-source PID seeks to decompose the mutual information \I{\source{0}\source{1} : \target} between ``sources'' \source{0} and \source{1} and a ``target'' \target into four nonnegative components.
The components identify information that is redundant, uniquely associated with \source{0}, uniquely associated with \source{1}, and synergistic:
\begin{align}
  \I{\source{0}\source{1} : \target} =
    & \phantom{+}~\Ipart{\source{0}\sep\source{1}}
	  & \textrm{\emph{redundant}} \nonumber \\
    & + \Ipart[\target \setminus \source{1}]{\source{0}}
	  & \textrm{\emph{unique with \source{0}}} \nonumber \\
    & + \Ipart[\target \setminus \source{0}]{\source{1}}
	  & \textrm{\emph{unique with \source{1}}} \nonumber \\
    & + \Ipart{\source{0}\source{1}} ~.
    & \textrm{\emph{synergistic}}
  \label{eq:decomp}
\end{align}
Furthermore, the mutual information $\I{\source{0} : \target}$ between \source{0} and \target is decomposed into two components:
\begin{align}
  \I{\source{0} : \target} =
    & \phantom{+}~\Ipart{\source{0}\sep\source{1}}
	  & \textrm{\emph{redundant}} \nonumber \\
    & + \Ipart[\target \setminus \source{1}]{\source{0}}
    ~.
    & \textrm{\emph{unique with \source{0}}}
  \label{eq:decompa}
\end{align}
And, similarly:
\begin{align}
  \I{\source{1} : \target} =
    & \phantom{+}~\Ipart{\source{0}\sep\source{1}}
    & \textrm{\emph{redundant}} \nonumber \\
    & + \Ipart[\target \setminus \source{0}]{\source{1}}
	  ~.
    & \textrm{\emph{unique with \source{1}}}
  \label{eq:decompb}
\end{align}
In this way, PID relates the four component informations.
However, since Eqs.\nobreakspace  \textup {(\ref {eq:decomp})} to\nobreakspace  \textup {(\ref {eq:decompb})}  provide only three independent constraints for four quantities, it does not uniquely determine how to quantify them in general.
That is, this fourth constraint lies outside of the PID.

By the same logic, though, the decomposition is uniquely determined by quantifying exactly one of its constituents.
In the case that one wishes to directly quantify the unique informations \Ipart[\target \setminus \source{1}]{\source{0}} and \Ipart[\target \setminus \source{0}]{\source{1}}, a consistency relation must hold so that they do not overconstrain the decomposition:
\begin{align}
  \Ipart[\target \setminus \source{1}]{\source{0}} & + \I{\source{1} : \target} \nonumber \\
  & = \Ipart[\target \setminus \source{0}]{\source{1}} + \I{\source{0} : \target}
  ~.
  \label{eq:consistency}
\end{align}
This ensures that using either Eq.\nobreakspace \textup {(\ref {eq:decompa})} or Eq.\nobreakspace \textup {(\ref {eq:decompb})} results in the same quantification of \Ipart{\source{0}\sep\source{1}}.

\section{Secret Key Agreement}
\label{sec:secrets}

\emph{Secret key agreement} is a fundamental concept within information-theoretic cryptography~\cite{maurer1993secret}.
Consider three parties---Alice, Bob, and Eve---who each partially observe a source of common randomness, joint probability distribution $ABE \sim p(a, b, e)$, where Alice has access only to $a$, Bob $b$, and Eve $e$.
The central challenge is to determine if it is possible for Alice and Bob to agree upon a secret key of which Eve has no knowledge.
The degree to which they may generate such a secret key immediately depends upon the structure of the joint distribution $ABE$.
It also depends upon whether Alice and Bob are allowed to publicly communicate.

Concretely, consider Alice, Bob, and Eve each receiving $n$ independent, identically distributed samples from $ABE$---Alice receiving $A^n$, Bob $B^n$, and Eve $E^n$.
A \emph{secret key agreement scheme} consists of functions $f$ and $g$, as well as a protocol ($h$) for public communication allowing either Alice, Bob, neither, or both to communicate.
In the case of a single party being permitted to communicate---say, Alice---she constructs $C = h(A^n)$ and then broadcasts it to all parties.
In the case that both parties are permitted communication, they take turns constructing and broadcasting messages of the form $C_i = h_i(A^n, C_{[0 \ldots i-1]})$ (Alice) and $C_i = h_i(B^n, C_{[0 \ldots i-1]})$ (Bob)~\cite{gohari2017coding}.

Formally, a secret key agreement scheme is considered $R$-achievable if for all $\epsilon > 0$:
\begin{align*}
  K_A &\stackrel{(1)}{=} f(A^n, C) ~, \\
  K_B &\stackrel{(2)}{=} g(B^n, C) ~, \\
  p(K_A = K_B = K) &\stackrel{(3)}{\geq} 1 - \epsilon ~, \\
  \I{K : C E^n} &\stackrel{(4)}{\leq} \epsilon ~, ~\text{and}\\
  \frac{1}{n} \H{K} &\stackrel{(5)}{\geq} R - \epsilon
  ~,
\end{align*}
where $(1)$ and $(2)$ denote the method by which Alice and Bob construct their keys $K_A$ and $K_B$, respectively, $(3)$ states that their keys must agree with arbitrarily high probability, $(4)$ states that the information about the key which Eve---armed with both her private information $E^n$ as well as the public communication $C$---has access to be arbitrarily small, and $(5)$ states that the key consists of approximately $R$ bits per sample.

The greatest rate $R$ such that an achievable scheme exists is known as the \emph{secret key agreement rate}.
Notational variations indicate which parties are permitted to communicate.
In the case that Alice and Bob are not allowed to communicate, their rate of secret key agreement is denoted \SKARzero{A}{B}{E}.
When only Alice is allowed to communicate their secret key agreement rate is \SKARone{A}{B}{E} or, equivalently, \SKARonel{B}{A}{E}.
When both Alice and Bob are allowed to communicate, their secret key agreement rate is denoted \SKARtwo{A}{B}{E}.
In this, we modified the standard notation for secret key agreement rates to emphasize which party or parties communicate.

In the case of no communication, \SKARzero{A}{B}{E} is given by~\cite{chitambar2018conditional}:
\begin{align}
  \SKARzero{A}{B}{E} = \H{A \meet B | E}
  ~,
  \label{eq:skarzero}
\end{align}
where $X \meet Y$ denotes the G{\'a}cs-K{\"o}rner common random variable~\cite{gacs1973common}.
It is worth noting that this quantity does not vary continuously with the distribution and generically vanishes.

In the case of one-way communication, \SKARone{A}{B}{E} is given by~\cite{ahlswede1993common}:
\begin{align}
  \SKARone{A}{B}{E} = \max \left\{ \I{B : K | C} - \I{E : K | C} \right\}
  ~,
  \label{eq:skarone}
\end{align}
where the maximum is taken over all variables $C$ and $K$, such that the following Markov condition holds: $C \markov K \markov A \markov BE$.
It suffices to consider $K$ and $C$ such that $|K| \leq |A|$ and $|C| \leq |A|^2$.

There is no such solution for \SKARtwo{A}{B}{E}; however, various upper- and lower-bounds are known~\cite{gohari2017coding}.
One simple lower bound is the supremum of the two one-way secret key agreement rates, as they are both a subset of bidirectional communication.
An even simpler upper bound that we will use is the \emph{intrinsic mutual information}~\cite{maurer1999unconditionally}:
\begin{align}
  \I{A : B \downarrow E} = \min_{p(\overline{e}|e)} \I{A : B | \overline{E}}
  ~.
  \label{eq:imi}
\end{align}
This effectively states that any information Eve has access to through any local modification of her observations cannot be secret.

The unique PID component \Ipart{\source{0} \setminus \source{1}} could be assigned the value of a secret key agreement rate under four different schemes.
First, neither \source{0} nor \target may be allowed to communicate.
Second, only \source{0} can communicate.
Third, only \target is permitted to communicate.
Finally, both \source{0} and \target may be allowed to communicate.
Note that the eavesdropper \source{1} is not allowed to communicate in any secret sharing schemes here.

Secret key agreement rates have been associated with unique informations before.
One particular upper bound on \SKARtwo{A}{B}{E}---the intrinsic mutual information~Eq.\nobreakspace \textup {(\ref {eq:imi})}---is known to not satisfy the consistency condition Eq.\nobreakspace \textup {(\ref {eq:consistency})}~\cite{bertschinger2013shared}.
More recently, the relationship between a particular method of quantifying unique information and one-way secret key agreement \SKARonel{\source{0}}{\target}{\source{1}}has been considered~\cite{banerjee2018unique}.

\section{Cryptographic Partial Information Decompositions}
\label{sec:decompositions}

\begin{figure}
  \includegraphics{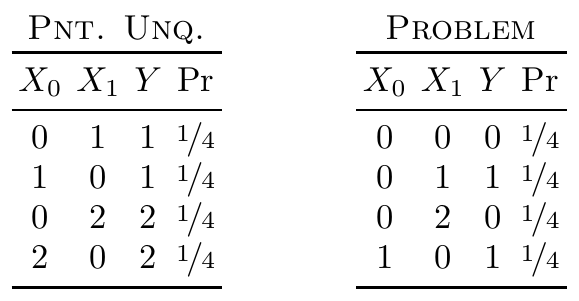}
  \caption{
    Two distributions of interest:
    The first, \texttt{Pointwise Unique}, exemplifies the directionality
	inherent in the one-way secret key agreement rates.
    The second, \texttt{Problem}, demonstrates that the no-communication,
	one-way communication with the source communicating (``camel''), and the
	two-way communication secret key agreement rates result in inconsistent
	PIDs.
  }
  \label{fig:distributions}
\end{figure}

We now address the application of each form of secret key agreement rate as unique information in turn.
For each resulting PID, we consider two distributions.
The first is that called \textsc{Pointwise Unique}, chosen here to exemplify the differing intuitions that can be applied to the PID.
The second distribution we look at is entitled \textsc{Problem} as it serves as a counterexample demonstrating that three of the four forms of secret key agreement do not result in a consistent decomposition.
Both distributions are given in Fig.\nobreakspace \ref {fig:distributions}.

Interpreting the \textsc{Pointwise Unique}~\cite{finn2018pointwise} distribution is relatively straightforward.
The target \target takes on the values `1' and `2' with equal probability.
At the same time, exactly one of the two sources (again with equal probability) will be equal to \target, while the other is `0'.
The mutual informations $\I{\source{0} : \target} = \SI{1/2}{\bit}$ and $\I{\source{1} : \target} = \SI{1/2}{\bit}$.

The \textsc{Problem} distribution lacks the symmetry of \textsc{Pointwise Unique}, yet still consists of four equally probable events.
The sources are restricted to take on pairs `00', `01', `02', `10'.
The target \target is equal to a `1' if either \source{0} or \source{1} is `1', and is `0' otherwise.
With this distribution, the mutual informations $\I{\source{0} : \target} = \SI{0.3113}{\bit}$ and $\I{\source{1} : \target} = \SI{1/2}{\bit}$.

\subsection{No Public Communication}
\label{subsec:zero}

In the first case, we consider the unique information from \source{i} to \target as the rate at which \source{i} and \target can agree upon a secret key while exchanging no public communication: $\Ipart{\source{i} \setminus \source{j}}~=~\SKARzero{\source{i}}{\target}{\source{j}}$.
This approach has some appeal: the PID is defined simply by a joint distribution without any express allowance or prohibition on public communication.
However, given its quantification in terms of the G{\'a}cs-K{\"o}rner common information, the quantity \SKARzero{\source{i}}{\target}{\source{j}} does not vary continuously with the distribution of interest.
Now, what is the behavior of this measure on our two distributions of interest?

When applied to \textsc{Pointwise Unique}, each source and the target are unable to construct a secret key.
In turn, each unique information is determined to be \SI{0}{\bit}.
This results in a redundancy and a synergy each of \SI{1/2}{\bit}.

The \textsc{Problem} distribution demonstrates the inability of \SKARzero{\source{i}}{\target}{\source{j}} to construct a consistent PID.
In this instance, as in the case of \textsc{Pointwise Unique}, no secrecy is possible and each unique information is assigned a value of \SI{0}{\bit}.
We therefore determine from Eq.\nobreakspace \textup {(\ref {eq:decompa})} that the redundancy should be $\I{\source{0} : \target} - \Ipart{\source{0} \setminus \source{1}} = \SI{0.3113}{\bit} - \SI{0}{\bit} = \SI{0.3113}{\bit}$.
Equation\nobreakspace \textup {(\ref {eq:decompb})}, however, says the redundancy is $\I{\source{1} : \target} - \Ipart{\source{1} \setminus \source{0}} = \SI{1/2}{\bit} - \SI{0}{\bit} = \SI{1/2}{\bit}$.
This contradiction demonstrates that no-communication secret key agreement rate cannot be used as a PID's unique components.

The resulting partial information decompositions for both distributions are listed in Table\nobreakspace \ref {tab:intuitionsa}.

\begin{table}
  \begin{tabular}{clr}
    \multicolumn{3}{c}{\SKARzero{\source{i}}{\target}{\source{j}}} \\
    \toprule
    \parbox[t]{2mm}{\multirow{4}{*}{\rotatebox[origin=c]{90}{\small\textsc{Pnt. Unq.}}}}
    & \Ipart{\source{0}\source{1}}                     & \SI{1/2}{\bit} \\
    & \Ipart[\target \setminus \source{1}]{\source{0}} & \SI{0}{\bit}   \\
    & \Ipart[\target \setminus \source{0}]{\source{1}} & \SI{0}{\bit}   \\
    & \Ipart{\source{0}\sep\source{1}}                 & \SI{1/2}{\bit} \\
    \midrule
    \parbox[t]{2mm}{\multirow{4}{*}{\rotatebox[origin=c]{90}{\small\textsc{Problem}}}}
    & \Ipart{\source{0}\source{1}}                     & \xmark       \\
    & \Ipart[\target \setminus \source{1}]{\source{0}} & \SI{0}{\bit} \\
    & \Ipart[\target \setminus \source{0}]{\source{1}} & \SI{0}{\bit} \\
    & \Ipart{\source{0}\sep\source{1}}                 & \xmark       \\
    \bottomrule
  \end{tabular}
  \caption{
    Partial information decompositions of \textsc{Pointwise Unique} and \textsc{Problem} when quantified using no-communication secret key agreement rate.
    \textsc{Pointwise Unique} decomposes into \SI{0}{\bit} for either unique
	information and into \SI{1/2}{\bit} for both the redundancy and synergy.
	\textsc{Problem}'s redundancy and synergy cannot be quantified, since the two secret key agreement rates result in different quantifications.
  }
\label{tab:intuitionsa}
\end{table}

\subsection{One-Way Public Communication}
\label{subsec:one}

We next consider the situation when one of the two parties is allowed public communication.
This gives us two options: either the source \source{i} communicates to target \target or \emph{vice versa}.
Both situations enshrine a particular \emph{directionality} in the resulting PID.

The first, where \source{i} constructs $C = h(\source{i}^n)$ and publicly communicates it, emphasizes the channels $\source{i} \to \target$ and creates a narrative of the sources conspiring to create the target.
We call this interpretation the \emph{camel intuition}, after the aphorism that a camel is a horse designed by committee.
The committee member \source{i} may announce what design constraints they brought to the table.

The second option, where \target constructs $C = h(\target^n)$ and publicly communicates it, emphasizes the channels $\target \to \source{i}$ and implies the situation that the sources are imperfect representations of the target.
We call this interpretation the \emph{elephant intuition}, as it recalls the parable of the blind men describing an elephant for the first time.
The elephant \target may announce which of its features is revealed in a particular instance.

\subsubsection{Camels}
\label{subsubsec:source}

The first option adopts $\Ipart{\source{i} \setminus \source{j}} = \SKARone{\source{i}}{\target}{\source{j}}$, bringing to mind the idea of sources acting as inputs into some scheme by which the target is produced.
When viewed this way, one may ask questions such as ``How much information in \source{0} is uniquely conveyed to \target?''.
Furthermore, the channels $\source{0} \to \target$ and $\source{1} \to \target$ take center stage.

Through this lens, the \textsc{Pointwise Unique} distribution has a clear interpretation.
Given any realization, exactly one source is perfectly correlated with the target, while the other is impotently `0'.
From this vantage, it is clear that the unique informations should each be \SI{1/2}{\bit}, and this is borne out with the one-way secret key agreement rate.
This implies that the redundancy and synergy of this decomposition are both
\SI{0}{\bit}.

For the \textsc{Problem} distribution, we find that \source{1} can broadcast the times when they observed a `1' or a `2', which correspond to \target having observed a `1' or `0', respectively.
In both instances \source{0} observed a `0' and so cannot deduce what the other two have agreed upon.
This leads to \SKARoner{\source{1}}{\target}{\source{0}} being equal to \SI{1/2}{\bit}.
At the same time, \SKARoner{\source{0}}{\target}{\source{1}} vanishes.
However, \textsc{Problem}'s redundancy and synergy cannot be quantified, since the two secret key agreement schemes imply different redundancies and so are inconsistent with Eq.\nobreakspace \textup {(\ref {eq:consistency})}.

The resulting PIDs for both are given in Table\nobreakspace \ref {tab:intuitionsb}.

\begin{table}
  \begin{tabular}{clr}
    \multicolumn{3}{c}{\SKARone{\source{i}}{\target}{\source{j}}} \\
    \toprule
    \parbox[t]{2mm}{\multirow{4}{*}{\rotatebox[origin=c]{90}{\small\textsc{Pnt. Unq.}}}}
    & \Ipart{\source{0}\source{1}}                     & \SI{0}{\bit}   \\
    & \Ipart[\target \setminus \source{1}]{\source{0}} & \SI{1/2}{\bit} \\
    & \Ipart[\target \setminus \source{0}]{\source{1}} & \SI{1/2}{\bit} \\
    & \Ipart{\source{0}\sep\source{1}}                 & \SI{0}{\bit}   \\
    \midrule
    \parbox[t]{2mm}{\multirow{4}{*}{\rotatebox[origin=c]{90}{\small\textsc{Problem}}}}
    & \Ipart{\source{0}\source{1}}                     & \xmark         \\
    & \Ipart[\target \setminus \source{1}]{\source{0}} & \SI{0}{\bit}   \\
    & \Ipart[\target \setminus \source{0}]{\source{1}} & \SI{1/2}{\bit} \\
    & \Ipart{\source{0}\sep\source{1}}                 & \xmark         \\
    \bottomrule
  \end{tabular}
  \caption{
    Partial information decompositions of \textsc{Pointwise Unique} and \textsc{Problem} when quantified using one-way communication secret key agreement rate with the source permitted public communication:
    \textsc{Pointwise Unique} decomposes into \SI{1/2}{\bit} for either unique
	information and into \SI{0}{\bit} for both the redundancy and synergy.
	\textsc{Problem}'s redundancy and synergy cannot be quantified.
  }
\label{tab:intuitionsb}
\end{table}

\subsubsection{Elephants}
\label{subsubsec:target}

When the target \target is the one party permitted communication, one adopts $\Ipart{\source{i} \setminus \source{j}} = \SKARonel{\source{i}}{\target}{\source{j}}$ and we can interpret the sources as alternate views of the singular target.
Consider, for example, journalism where several sources give differing perspectives on the same event.
When viewed this way, one might ask a question such as ``How much information in \target is uniquely captured by \source{0}?''.
The channels $\source{0} \leftarrow \target$ and $\source{1} \leftarrow \target$ are paramount with this approach.
We denote these in reverse to emphasize that \target is still the \textit{target} in the PID.

Considered this way, the \textsc{Pointwise Unique} distribution takes on a different character.
The sources each receive identical descriptions of the target---accurate half the time and erased the remainder.
The description is identical, however.
Nothing is uniquely provided to either source.
This is reflected in the secret key agreement rates, which are \SI{0}{\bit}, leaving both the redundancy and synergy \SI{1/2}{\bit}.

The \textsc{Problem} distribution's unique informations are $\SKARonel{\source{0}}{\target}{\source{1}} = \SI{0}{\bit}$ and $\SKARonel{\source{1}}{\target}{\source{0}} = \SI{0.1887}{\bit}$.
Unlike the prior two decompositions, these unique informations satisfy Eq.\nobreakspace \textup {(\ref {eq:consistency})}.
The resulting redundancy is \SI{0.3113}{\bit} while the synergy is \SI{1/2}{\bit}.

Their PIDs are listed in Table\nobreakspace \ref {tab:intuitionsc}.
Thus, by having \target publicly communicate and so invoking a particular directionality we, finally, get a consistent PID.

\begin{table}
  \begin{tabular}{clr}
    \multicolumn{3}{c}{\SKARonel{\source{i}}{\target}{\source{j}}} \\
    \toprule
    \parbox[t]{2mm}{\multirow{4}{*}{\rotatebox[origin=c]{90}{\small\textsc{Pnt. Unq.}}}}
    & \Ipart{\source{0}\source{1}}                     & \SI{1/2}{\bit} \\
    & \Ipart[\target \setminus \source{1}]{\source{0}} & \SI{0}{\bit}   \\
    & \Ipart[\target \setminus \source{0}]{\source{1}} & \SI{0}{\bit}   \\
    & \Ipart{\source{0}\sep\source{1}}                 & \SI{1/2}{\bit} \\
    \midrule
    \parbox[t]{2mm}{\multirow{4}{*}{\rotatebox[origin=c]{90}{\small\textsc{Problem}}}}
    & \Ipart{\source{0}\source{1}}                     & \SI{1/2}{\bit}    \\
    & \Ipart[\target \setminus \source{1}]{\source{0}} & \SI{0}{\bit}      \\
    & \Ipart[\target \setminus \source{0}]{\source{1}} & \SI{0.1887}{\bit} \\
    & \Ipart{\source{0}\sep\source{1}}                 & \SI{0.3113}{\bit} \\
    \bottomrule
  \end{tabular}
  \caption{
    PID for \textsc{Pointwise Unique} and \textsc{Problem} when quantified using one-way communication secret key agreement rate with the target permitted public communication:
    \textsc{Pointwise Unique} decomposes into \SI{1/2}{\bit} for either unique information, and into \SI{0}{\bit} for both the redundancy and synergy.
    \textsc{Problem} admits unique informations of \SI{0}{\bit} and \SI{0.1887}{\bit}, respectively.
    This results in a redundancy of \SI{0.3113}{\bit} and a synergy of \SI{1/2}{\bit}, providing a consistent PID.
  }
\label{tab:intuitionsc}
\end{table}

\subsection{Two-Way Public Communication}
\label{subsec:two}

We finally turn to the full two-way secret key agreement rate: $\Ipart{\source{i} \setminus \source{j}} = \SKARtwo{\source{i}}{\target}{\source{j}}$.
This approach is also appealing, as it does not ascribe any directionality to interpreting the PID.
Furthermore, it varies continuously with the distribution, unlike the no-communication case.
However, this quantity is generally impossible to compute directly, with only upper and lower bounds known.
Fortunately, this only slightly complicates the analyses we wish to make.

In the case of the \textsc{Pointwise Unique} distribution, it is not possible to extract more secret information than was done in the camel situation.
Therefore, the resulting PID is identical: unique informations of \SI{1/2}{\bit} and redundancy and synergy of \SI{0}{\bit}.

\textsc{Problem}, however, is again a problem.
Upper and lower bounds on \SKARtwo{\source{1}}{\target}{\source{0}} converge\footnote{In this instance, the larger of the two one-way secret key agreement rates form a lower bound of \SI{1/2}{\bit}. While the upper bound provided by the intrinsic mutual information is also \SI{1/2}{\bit}.} to \SI{1/2}{\bit}, and so we know this value exactly.
Utilizing the consistency relation Eq.\nobreakspace \textup {(\ref {eq:consistency})}, we find that the other unique information must be \SI{0.3113}{\bit} in order for the full decomposition to be consistent.
However, the intrinsic mutual information places an upper bound of \SI{0.1887}{\bit} on \SKARtwo{\source{0}}{\target}{\source{1}}.
We therefore must conclude that two-way secret key agreement rates cannot be used to directly quantify unique information and a consistent PID cannot be built using them.

The resulting PIDs for both these distributions can be seen in Table\nobreakspace \ref {tab:intuitionsd}.

\begin{table}
  \begin{tabular}{clr}
    \multicolumn{3}{c}{\SKARtwo{\source{i}}{\target}{\source{j}}} \\
    \toprule
    \parbox[t]{2mm}{\multirow{4}{*}{\rotatebox[origin=c]{90}{\small\textsc{Pnt. Unq.}}}}
    & \Ipart{\source{0}\source{1}}                     & \SI{0}{\bit}   \\
    & \Ipart[\target \setminus \source{1}]{\source{0}} & \SI{1/2}{\bit} \\
    & \Ipart[\target \setminus \source{0}]{\source{1}} & \SI{1/2}{\bit} \\
    & \Ipart{\source{0}\sep\source{1}}                 & \SI{0}{\bit}   \\
    \midrule
    \parbox[t]{2mm}{\multirow{4}{*}{\rotatebox[origin=c]{90}{\small\textsc{Problem}}}}
    & \Ipart{\source{0}\source{1}}                     & \xmark                   \\
    & \Ipart[\target \setminus \source{1}]{\source{0}} & $\leq$ \SI{0.1887}{\bit} \\
    & \Ipart[\target \setminus \source{0}]{\source{1}} & \SI{1/2}{\bit}           \\
    & \Ipart{\source{0}\sep\source{1}}                 & \xmark                   \\
    \bottomrule
  \end{tabular}
  \caption{
    PID for \textsc{Pointwise Unique} and \textsc{Problem} when quantified
	using two-way communication secret key agreement rate:
    \textsc{Pointwise Unique} decomposes into \SI{1/2}{\bit} for either unique
	information, and into \SI{0}{\bit} for both the redundancy and synergy.
	\textsc{Problem}'s redundancy and synergy cannot be quantified, because the two secret key agreement rates result in different quantifications.
  }
\label{tab:intuitionsd}
\end{table}

\subsection{Summary}
\label{sec:PID_SKAR_summary}

To conclude, then, there is only one secret-key communication scenario---\target publicly communicates---that yields a consistent PID, as in Table\nobreakspace \ref {tab:intuitionsc}.
While we have not proven this, we have been unable to find a counterexample after extensive numerical searches using the \texttt{dit}~\cite{dit} software package.
That is, one must invoke a directionality, unspecified by the PID, to have a consistent PID if using secret key agreement as the basis for the PID component of unique information.

\section{Discussion}
\label{sec:discussion}

We now turn to two follow-on developments arising from the tools developed thus far.
First, we define a distribution whose two-way secret key agreement rates behave in a curious manner with very interesting implications regarding the nature of information itself.
Second, we take a closer look at an alternative proposal for quantifying unique information and describe its behavior in relationship to the camel/elephant dichotomy defined in Section\nobreakspace \ref {sec:decompositions}.

\subsection{When Conversation is More Powerful Than a Lecture}
\label{subsec:twowaydist}

\begin{figure}
  \includegraphics{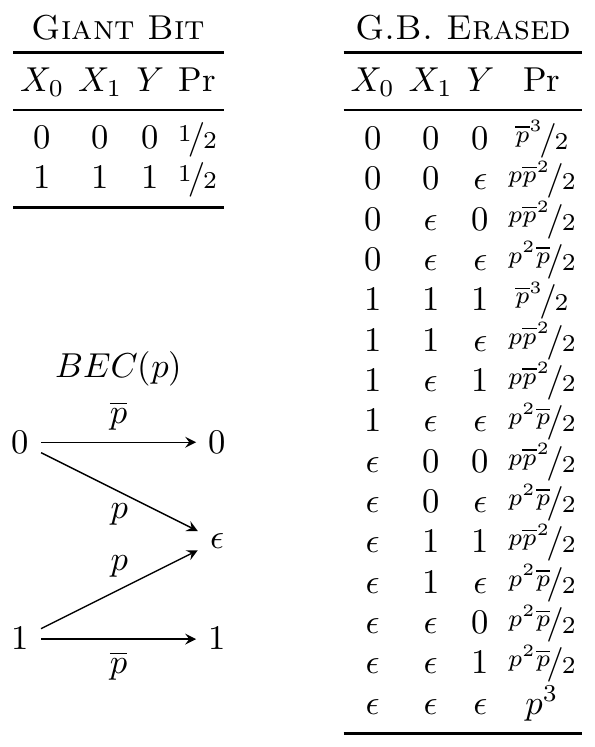}
  \caption{
    Distribution whose one-way secret key agreement rates are all \SI{0}{\bit}, yet has nonzero two-way secret key agreement rate.
    It is constructed from the \textsc{Giant Bit} distribution by
	passing each variable independently through a binary erasure channel
	$BEC(p)$ with erasure probability $p$.
    This distribution has a two-way secret key agreement rate of $p\pbar^2$ between any two variables with the third as an eavesdropper.
  }
  \label{fig:moredistributions}
\end{figure}

We now explore the PID quantified by two-way secret key agreement further.
Consider the \textsc{Giant Bit} distribution, which exemplifies redundant information.
The distribution \textsc{G.B. Erased}, resulting from passing each variable through an independent binary erasure channel (BEC), exhibits many interesting properties.
It is listed in Fig.\nobreakspace \ref {fig:moredistributions}.
Most notably, the one-way secret key agreement rates between any two variables with the third eavesdropping vanish.
However, the two-way secret key agreement rate is equal to $p\pbar^2 = \I{\source{i} : \target | \source{j}}$~\cite{gohari2017comments}.
Furthermore, notice that subtracting Eq.\nobreakspace \textup {(\ref {eq:decompb})} from Eq.\nobreakspace \textup {(\ref {eq:decomp})} tells us that:
\begin{align}
  \I{\source{0}\source{1} : \target} & - \I{\source{1} : \target} \nonumber \\
    &= \I{\source{0} : \target | \source{1}} \nonumber \\
    &= \Ipart[\target \setminus \source{1}]{\source{0}} + \Ipart{\source{0}\source{1}}
    .
\end{align}
That is, the conditional mutual information is equal to unique information plus synergistic information.

Evaluating the PID using \SKARtwo{\source{i}}{\target}{\source{j}} as unique information results, in this case, in a consistent decomposition.
Furthermore, the redundant and synergistic informations are zero.
This is, however, troublesome: \textsc{G.B. Erased} possesses nonzero third-order connected information~\cite{Schneidman2003}, a quantity commonly considered a component of the synergy~\cite{james2017unique}.
Indeed, it is provably attributed to synergy by both the \Idep[]{}~\cite{james2017unique} and \Ibroja[]{}~\cite{bertschinger2014quantifying} methods, and likely others as well.
No other proposed method of quantifying the PID results in zero redundancy or synergy.
The implication here is that, if indeed the third-order connected information is a component of synergy, the two-way secret key agreement rate \emph{overestimates} unique information by including some types of synergistic effect.
Therefore, we conclude that bidirectional communication between two parties can, in some instances, determine information held solely in trivariate interactions.
It remains to be understood (i) how independently and identically transforming a distribution with no third-order connected information can result in its creation and (ii) how only two of the variables can recover it when allowed to communicate.

\subsection{\Ibroja[]{}, the Elephant}
\label{subsec:ibroja}

The measure of Bertschinger \etal~\cite{bertschinger2014quantifying}, here referred to as \Ibroja[]{}, is perhaps the most widely accepted and used method of quantifying the PID.
Though popular, it has its detractors~\cite{ince2017measuring,finn2018pointwise}.
Here, we interpret the criticisms leveled and \Ibroja[]{} as a product of camel intuitions being applied to an elephantesque~\cite{banerjee2018unique} measure.
In doing so, we will primarily consider the \textsc{Pointwise Unique} distribution.

As noted in Section\nobreakspace \ref {subsec:one}, if a source is permitted to communicate with the target, then a secret key agreement rate of \SI{1/2}{\bit} is achievable; while if the target communicates with the source then it is impossible to agree upon a secret key.
From this camel perspective it is clear that each source, half the time, uniquely determines the target.
The elephant perspective, however, allots nothing to unique informations as each source is provided with identical information.
This would greatly disconcert the camel and may lead one to think that the elephant has ``artificially inflated'' the redundancy.
We next take a closer look at this notion, using \Ibroja[]{}.

\begin{figure}
  \includegraphics{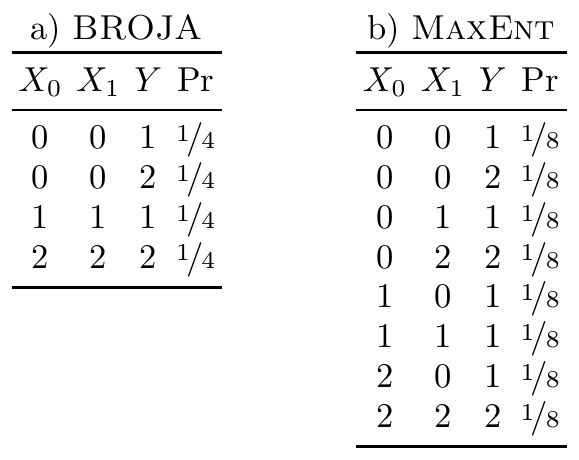}
  \caption{
    Two modified forms of the \textsc{Pointwise Unique} distribution.
    a) Intermediate distribution resulting from the BROJA optimization. It has
	the minimum sources-target mutual information consistent with the
	source-target marginals.
    b) Maximum entropy distribution consistent with the source-target marginals.
    It contains no structure beyond that implied by those marginals.
  }
  \label{fig:pntunqdists}
\end{figure}

In the course of computing \Ibroja[]{} for the distribution $p(\source{0}, \source{1}, \target)$, the set of distributions:
\begin{align*}
  Q = \left\{ q(\source{0}, \source{1}, \target) : \forall i,~ q(\source{i}, \target) = p(\source{i}, \target) \right\}
\end{align*}
is considered.
The \assumption assumption~\cite{bertschinger2014quantifying} is then invoked, which states that redundancy and all unique informations are constant within this family of distributions.
To complete the quantification, the distribution with minimum \I{\source{0}\source{1} : \target} is selected from this family.
The resulting distribution associated with the \textsc{Pointwise Unique} distribution can be seen in Fig.\nobreakspace \ref {fig:pntunqdists}.
Made explicit, it can now be seen that \Ibroja[]{} does indeed correlate the sources, but under assumption \assumption this does not effect the redundancy.

One aspect of \Ibroja[]{} and assumption \assumption we believe warrants further investigation is its relationship with \emph{maximum entropy} philosophy~\cite{Jayn83}.
The latter is, in effect, Occam's razor applied to probability distributions: given a set of constraints, the most natural distribution to associate with them is that with maximum entropy.
As it turns out, this is equivalent to the distribution nearest the unstructured product-of-marginals distribution $\overline{p}(x, y, z, \ldots) = p(x)p(y)p(z)\ldots$~\cite{amari2001information}:
\begin{align*}
  \argmax_{q \in Q} \H{q} = \argmin_{q \in Q} \DKL{q}{\overline{p}}
  ~,
\end{align*}
where $\DKL{P}{P}$ is the relative entropy between distributions $P$ and $Q$.
Having briefly introduced the ideas behind maximum entropy, we next cast their light on the BROJA optimization employed to calculate \Ibroja[]{}.

Let us first consider the distribution resulting from BROJA optimization.
Its entropy is unchanged from the \textsc{Pointwise Unique} distribution indicating that it has the same amount of structure---they are equally distant from the product distribution.
The BROJA distribution has a reduced \I{\source{0}\source{1} : \target} mutual information, however, indicating perhaps that the optimization has shifted some of the distribution's structure away from the sources-target interaction.
It is interesting that this optimization could not simply remove the synergy from the distribution altogether, resulting in a larger entropy.

If one takes assumption \assumption and directly applies the maximum entropy philosophy, a different distribution results.
This distribution, seen in Fig.\nobreakspace \ref {fig:pntunqdists}, has a larger entropy than both the \textsc{Pointwise Unique} and the BROJA intermediate distribution, indicating that it in fact has less structure than either.
Under assumption \assumption, the \textsc{MaxEnt} distribution, also in Fig.\nobreakspace \ref {fig:pntunqdists}, retains all the redundant and unique informations, while under maximum entropy it contains no structure not implied by the source-target marginals---\eg, no synergy.

To be clear, this is not to claim that assumption \assumption or BROJA optimization are \emph{wrong} or \emph{incorrect}, only that the optimization's behavior in light of well-established maximum entropy principles is subtle and requires a careful investigation.
For example, it may be that the source-target marginals do imply some level of triadic interaction and therefore the maximum entropy distribution reflects this lingering synergy.
At the same time, BROJA minimization may be capable of maintaining that level of structure implied by the marginals, but somehow shunts it into \H{\target | \source{0}\source{1}}.

\section{Conclusion}
\label{sec:conclusion}

At present, a primary barrier for PID's general adoption as a useful and possibly a central tool in analyzing how complex systems store and process information is an agreement on a method to quantify its component informations.
Here, we posited that one reason for disagreement stems from conflicting intuitions regarding the decomposition's operational behavior.
To give an operational meaning to unique information and address these intuitions, we equated unique information with the ability of two parties to agree upon a secret.
This leads to numerous observations.

The first is that the PID, as currently defined, is ambivalent to any notion of directionality.
There are, however, very clear cases in which the assumption of a directionality---or lack there of---is critical to the existence of unique information.
Consider, for example, the case of the McGurk effect~\cite{mcgurk1976hearing} where the visual stimulus of one phoneme and the auditory stimulus of another phoneme gives rise to the perception of a third phoneme.
By construction, the stimuli \emph{cause} the perception, and the channels implicit in a camel intuition are central.
If one were to study this interaction using an elephant-like PID, it is unclear that the resulting decomposition would reflect the neurobiological mechanisms by which the perception is produced.
Similarly, a camel-like measure would be inappropriate when interpreting simultaneous PET and MRI scans of a tumor.

One can view this as the PID being inherently context-dependent and conclude that quantification requires specifying directionality.
In this case, the elephant intuition is apparently more natural, as adopting closely-related notions from cryptography results in a consistent PID.
If context demands the camel intuition, though, either a noncryptographic method of quantifying unique information is needed or consistency must be enforced by augmenting the secret key agreement rate.
It is additionally possible that associating secret key agreement rates with unique information is fundamentally flawed and that, ultimately, PID entails quantifying unique information as something distinct from the ability to agree upon a secret key.
This missing thing has yet to be identified.

The next observation concerns the third-order connected information.
We first demonstrated that such triadic information can be constructed from common information from which each variable is then independently and identically modified.
Furthermore, it has been shown that two of those three parties, when engaging in bidirectional communication, can capture this triadic information.
This does not generically occur: For example, if \source{0}\source{1}\target are related by \textsc{Xor}, the distribution contains \SI{1}{\bit} of third-order connected information, but \SKARtwo{\source{0}}{\target}{\source{1}} (or any permutation of the variables) is equal to \SI{0}{\bit}.
This suggests that the third-order connected information may not be an atomic quantity, but rather consists of two parts, one accessible to two communicating parties and one not.

Our third observation regards the behavior of the \Ibroja[]{} measure, especially in relation to standard maximum entropy principles.
We first demonstrated that \Ibroja[]{} indeed correlates sources, but argued that this behavior only seems inappropriate when adopting a camel intuition.
We then discussed how its intermediate distribution is as structured as the initial one and so if indeed \Ibroja[]{} is operating correctly, it must shuffle the dependencies that result in synergy to another aspect of the distribution.
Finally, we discussed how the standard maximum entropy approach may remove synergy from a distribution all together.
This calls for a more careful investigation as to whether it does---and BROJA optimization is incorrect---or does not---and synergistic information is implied under source-target marginals and Occam's razor.

Looking to the future, we trust that this exploration of the relationship between cryptographic secrecy and unique information will provide a basis for future efforts to understand and quantify the partial information decomposition.
Furthermore, the explicit recognition of the role that directional intuitions
play in the meaning and interpretation of a decomposition should reduce
cross-talk and improve understanding as we collectively move forward.

\section*{Acknowledgments}
\label{sec:acknowledgments}

All calculations were performed using the \texttt{dit} Python package~\cite{dit}.
We thank P. Banerjee, E. Olbrich, S. Loomis, and D. Feldspar for many helpful discussions.
As a faculty member, JPC thanks the Santa Fe Institute and the Telluride Science Research Center for their hospitality during visits.
This material is based upon work supported by, or in part by, Foundational Questions Institute grant FQXi-RFP-1609, the U.S. Army Research Laboratory and the U.S. Army Research Office under contracts W911NF-13-1-0390 and W911NF-13-1-0340 and grant W911NF-18-1-0028, and via Intel Corporation support of CSC as an Intel Parallel Computing Center.


\begin{thebibliography}{10}

\bibitem{williams2010nonnegative}
P.~L. Williams and R.~D. Beer.
\newblock Nonnegative decomposition of multivariate information.
\newblock {\em arXiv:1004.2515}.

\bibitem{rauh2017extractable}
J.~Rauh, P.~Banerjee, E.~Olbrich, J.~Jost, and N.~Bertschinger.
\newblock On extractable shared information.
\newblock {\em Entropy}, 19(7):328, 2017.

\bibitem{rauh2017secret}
J.~Rauh.
\newblock Secret sharing and shared information.
\newblock {\em Entropy}, 19(11):601, 2017.

\bibitem{maurer1993secret}
U.~M. Maurer.
\newblock Secret key agreement by public discussion from common information.
\newblock {\em IEEE Trans. Info. Th.}, 39(3):733--742, 1993.

\bibitem{gohari2017coding}
A.~Gohari, O.~G{\"u}nl{\"u}, and G.~Kramer.
\newblock Coding for positive rate in the source model key agreement problem.
\newblock {\em arXiv:1709.05174}.

\bibitem{chitambar2018conditional}
E.~Chitambar, B.~Fortescue, and M.-H. Hsieh.
\newblock The conditional common information in classical and quantum secret
  key distillation.
\newblock {\em IEEE Trans. Info. Th.}, 2018.

\bibitem{gacs1973common}
P.~G{\'a}cs and J.~K{\"o}rner.
\newblock Common information is far less than mutual information.
\newblock {\em Problems of Control and Information Theory}, 2(2):149--162,
  1973.

\bibitem{ahlswede1993common}
R.~Ahlswede and I.~Csisz{\'a}r.
\newblock Common randomness in information theory and cryptography. {I. Secret}
  sharing.
\newblock {\em IEEE Trans. Info. Th.}, 39(4):1121--1132, 1993.

\bibitem{maurer1999unconditionally}
U.~M. Maurer and S.~Wolf.
\newblock Unconditionally secure key agreement and the intrinsic conditional
  information.
\newblock {\em IEEE Trans. Info. Th.}, 45(2):499--514, 1999.

\bibitem{bertschinger2013shared}
N.~Bertschinger, J.~Rauh, E.~Olbrich, and J.~Jost.
\newblock Shared information—new insights and problems in decomposing
  information in complex systems.
\newblock In {\em Proceedings of the European Conference on Complex Systems
  2012}, pages 251--269. Springer, 2013.

\bibitem{banerjee2018unique}
P.~K. Banerjee, E.~Olbrich, J.~Jost, and J.~Rauh.
\newblock Unique informations and deficiencies.
\newblock {\em arXiv:1807.05103}.

\bibitem{finn2018pointwise}
C.~Finn and J.~T. Lizier.
\newblock Pointwise partial information decomposition using the specificity and
  ambiguity lattices.
\newblock {\em Entropy}, 20(4):297, 2018.

\bibitem{dit}
R.~G. James, C.~J. Ellison, and J.~P. Crutchfield.
\newblock {dit}: a {P}ython package for discrete information theory.
\newblock {\em J. Open Source Software}, 3(25):738, 2018.

\bibitem{gohari2017comments}
A.~Gohari and V.~Anantharam.
\newblock Comments on ``information-theoretic key agreement of multiple
  terminals: {Part I}''.
\newblock {\em IEEE Trans. Info. Th.}, 63(8):5440--5442, 2017.

\bibitem{Schneidman2003}
E.~Schneidman, S.~Still, M.~J. Berry, W.~Bialek, et~al.
\newblock Network information and connected correlations.
\newblock {\em Phys. Rev. Lett.}, 91(23):238701, 2003.

\bibitem{james2017unique}
R.~G. James, J.~Emenheiser, and J.~P. Crutchfield.
\newblock Unique information via dependency constraints.
\newblock {\em J. Phys. A}, in press, 2018.
\newblock arXiv:1709.06653.

\bibitem{bertschinger2014quantifying}
N.~Bertschinger, J.~Rauh, E.~Olbrich, J.~Jost, and N.~Ay.
\newblock Quantifying unique information.
\newblock {\em Entropy}, 16(4):2161--2183, 2014.

\bibitem{ince2017measuring}
R.~A.A. Ince.
\newblock Measuring multivariate redundant information with pointwise common
  change in surprisal.
\newblock {\em Entropy}, 19(7):318, 2017.

\bibitem{Jayn83}
E.~T. Jaynes.
\newblock Where do we stand on maximum entropy?
\newblock In E.~T. Jaynes, editor, {\em Essays on Probability, Statistics, and
  Statistical Physics}, page 210. Reidel, London, 1983.

\bibitem{amari2001information}
S.~Amari.
\newblock Information geometry on hierarchy of probability distributions.
\newblock {\em IEEE Trans. Info. Th.}, 47(5):1701--1711, 2001.

\bibitem{mcgurk1976hearing}
H.~McGurk and J.~MacDonald.
\newblock Hearing lips and seeing voices.
\newblock {\em Nature}, 264(5588):746, 1976.

\end{thebibliography}
\end{document}